\documentclass[intlimits,twoside,a4paper]{article}

\usepackage[cp1251]{inputenc}
\usepackage{xcolor}
\usepackage[eqsecnum]{cmpj3}

\usepackage{bm}

\issue{2021}{24}{1}{13704}
\doinumber{10.5488/CMP.24.13704}

	\title[Anisotropic Heisenberg model for the mixed spin-3/2 and spin-1/2 under random crystal field]
		{Anisotropic Heisenberg model for the mixed spin-3/2 and spin-1/2 under random crystal field}
	
\author[D. Sabi Takou, M. Karimou, F. Hontinfinde, E. Albayrak]
{D. Sabi Takou\refaddr{label1,label3}, M. Karimou\refaddr{label2,label3}, F. Hontinfinde\refaddr{label3,label4},  
	E. Albayrak\refaddr{label5}}
		
		\addresses{
		\addr{label1} Ecole Polytechnique d'Abomey-Calavi (EPAC-UAC), Department of  Fundamentale Sciences, Republic of Benin
		\addr{label2} Ecole Nationale Sup\'{e}rieure de G\'{e}nie Energ\'{e}tique et Proc\'{e}d\'{e}s (ENSGEP) d'Abomey, Republic of Benin
		\addr{label3} Institute of Mathematics and Physical Sciences (IMSP), Republic of Benin
		\addr{label4} University of Abomey-Calavi, Department of Physics, Republic of Benin
		\addr{label5} Erciyes University, Department of Physics, 38039, Kayseri, Turkey}
		
\date{Received July 24, 2020, in final form September 28, 2020}
		
\begin{document}
		
		\maketitle%
	\begin{abstract}

 Thermodynamic properties of the mixed spin-3/2 and spin-1/2
	Heisenberg model are examined within the Oguchi approximation
in the presence of a random crystal-field (RCF).
	The RCF is either introduced with probability $p$ or turned off with
	probability $1-p$ randomly. The thermal variations of the global magnetization
	 and free energy of the system are investigated to construct the phase diagrams
	 for the  classical, quantum and anisotropic cases. Different results
        revealed that no qualitative changes exist between them. Quantum effects are found to be present
	and abundant in the quantum model in the negative $ D $-range. This phenomenon
	 has a strong decreasing effect on the critical temperature which becomes much lower than
	  in the classical case. In the presence of an external
	 field, it was observed that coercivity and remanence decrease in a wide range of
	 the absolute temperature.

\keywords exchange anisotropy, Heisenberg model, Oguchi approximation, mixed spin,  crystal field
\end{abstract}

\section{Introduction}

In the classical spin models, e.g., Ising model (IM), the spins can only lie along one
chosen axis which is often taken to be the $z$-axis. These systems may be
relatively easy to deal with when compared with the quantum spin models, i.e., Heisenberg
model (HM), because the spins now have freedom to orient themselves in any
three-dimensional space. This, of course, leads to an uncertainty since the spin
operators, i.e., the components of a spin, do not commute with each other. Therefore,
it is usually impossible to obtain an analytical solution. Their solutions often
require some possible approximations which may provide qualitative pictures
but mostly presents some shortcomings. 

The Oguchi approximation (OA) is such an approximation which is used
to solve the HMs. The spin-1/2 anistropic HM was studied with Dzyaloshinsky-Moriya
interaction \cite{Sousa}. The effects of the second-nearest-neighbor exchange
interactions on the magnetization, internal energy, heat capacity, entropy and free energy
were examined \cite{Mert}. The phase diagrams of the mixed spin-1 and spin-1/2
under the influence of both exchange and single-ion anisotropies were taken
into consideration \cite{Bobak}. This model was also extended to the studies
of compensation temperature \cite{Bobak1} and magnetic susceptibility
\cite{Bobak2}. The mixed spin-2 and spin-1/2 HM was studied under the
influence of exchange anisotropy and crystal field by using the OA on
the square \cite{Albayrak0} and simple cubic lattices \cite{Albayrak1}.
A similar study  was also performed on the ferromagnetic mixed
spin-3/2 and spin-1/2 model for the HM in terms of the OA \cite{Bobak3}. 

In addition to the HM in the OA approximation, there are some other works
which use some other techniques, considering  various aspects in the IM.
The first technique is the effective field theory (EFT) which was considered
in the calculation of magnetizations and phase diagrams \cite{Kaneyoshi0},
the critical properties in a transverse field \cite{Kaneyoshi2}, magnetic
properties in a longitudinal magnetic field \cite{Wei}, the magnetic and
hysteresis behaviors for a bilayer model \cite{Essaoudi}, bimodal random-crystal
field distribution effects \cite{Albayrakx}, for a ferromagnetic or antiferromagnetic
bilayer system with transverse field \cite{Jiang1,Saber}, in a random field \cite{Liang},
and for a trimodal random-field distribution \cite{Liang1}. Some of the works
with a diluted model were also examined in the EFT in a random field \cite{Kaneyoshi1},
with coordination numbers 3 and 4 \cite{Bobak0}, in a longitudinal random
field \cite{Liang2}, for transverse Ising model \cite{Benayad0,Jiang0} and
for the study of the magnetic properties\cite{Bobak1x}. In addition to the
EFT studies, the Monte Carlo algorithm \cite{xBuendia}, the exact recursion
relations (ERR) were used on the Bethe lattice (BL) \cite{xAlbayrak}, by establishing
a mapping correspondence with the eight-vertex model \cite{xStrecka}, 
an exact star-triangle mapping transformation \cite{xJascur}, the ERR's on 
a two-fold Cayley tree~\cite{xZhang}, on the Union Jack lattice by means of
 a mapping correspondence with the eight-vertex model
\cite{xStrecka0}, on a rope ladder it was examined by combining two exact
analytical methods, i.e., decoration-iteration mapping transformation and
standard transfer-matrix method \cite{xKissova}, within the mean-field
approximation~\cite{xBahmad0} and with the use of the ERR's on the BL
\cite{xAlbayrak2}, on a two-layer BL \cite{xAlbayrak3} and $\pm J$ model \cite{xAlbayrak4}. 

In this work, we extend the study of the  mixed-spin 3/2 and 1/2 HM in
the  OA \cite{Bobak3} with the inclusion of the RCF and examine its
phase diagrams by studying the thermal variations of total and relative
magnetizations as well as the free energy. The  hysteresis cycles behavior
is also analysed at different temperatures.

The rest of the work is arranged as follows: The next section 
is devoted to the formulation of Heisenberg model in the Oguchi
approximation for the mixed spin-3/2 and 1/2 system under the effect
of the RCF. The third section consists of illustrations and possible
discussions and comparisons. The final section is devoted to a brief summary and conclusion.  
  
\section{The formulation of the HM in the OA for the mixed-spin 3/2 and  1/2 model} 

 A lattice with coordination number $q$ is considered.
 It is devided into two equivalent sublattices A and B 
with spins 1/2  and 3/2, respectively.  The usual Heisenberg Hamiltonian is considered.
 It includes the nearest-neighbor (NN) bilinear interaction parameter $J$,
 the random crystal field $D_i$ active only at spin-3/2 sites
 and the external magnetic field $h$. Its expression follows:  

%
\begin{eqnarray}\label{eq:Hamiltonian}
\hat{\mathcal{H}} = &-& J \sum\limits_{\langle i j
 \rangle}\left[(1-\Delta)( \hat{S}_{i,3/2}^x \hat{S}_{j,1/2}^x+\hat{S}_{i,3/2}^y
\hat{S}_{j,1/2}^y)+\hat{S}_{i,3/2}^z \hat{S}_{j,1/2}^z\right] \nonumber\\
&-&  \sum\limits_{i} D_i {\left(\hat{S}_{i,3/2}^z\right)^2}-h\left(\sum_{i}S_{i,3/2}^{z} +
\sum_{j}S_{j,1/2}^{z}\right),
\end{eqnarray}
where $S_{i,3/2}^\nu$ and $S_{j,1/2}^\nu$ with $\nu=x,y,z$ are the components of spin operators for spin-3/2 and spin-1/2 and $\Delta$ is the exchange anisotropy parameter with $0\leqslant\Delta\leqslant1$. The values of $\Delta$ changes it into the IM Hamiltonian when it is set to one, into the HM Hamiltonian when it is equal to zero and into the anisotropic model (AM) Hamiltonian when $0<\Delta<1$.  

To solve the model, some approximations are necessary.
In the OA,  the effective Oguchi Hamiltonian for a pair of 3/2 and 1/2 spins is given as   
\begin{eqnarray}
\hat{\mathcal{H}}_{ij} = &-& J \left[ (1-\Delta)( \hat{S}_{i,3/2}^x \hat{S}_{j,1/2}^x+\hat{S}_{i,3/2}^y \hat{S}_{j,1/2}^y)+\hat{S}_{i,3/2}^z \hat{S}_{j,1/2}^z\right]  \nonumber\\
&-& D_i {\left(\hat{S}_{i,3/2}^z\right)^2}-h_i \hat{S}_{i,3/2}^z- h_j \hat{S}_{j,1/2}^z\,,
\end{eqnarray}
where the mean-field terms are given as
\begin{eqnarray}
h_i=h+(q-1) J M_{1/2}\,, \nonumber \\ h_j=h+(q-1) J M_{3/2}\,,
\end{eqnarray}
with $M_{3/2}=\langle \hat{S}_{i,3/2}^z \rangle$ and $M_{1/2}=\langle \hat{S}_{j,1/2}^z \rangle$ being the magnetizations of the sublattices with spin-3/2 and spin-1/2, $q$ is the number of NN's, and $\langle... \rangle$ corresponds to thermal averages.

One has to solve the eigenvalue equation
\begin{equation}\label{eq:Eigen}
\hat{\mathcal{H}}_{ij} \mid n\rangle=H_n \mid n\rangle, \qquad    n=1,2,\ldots,8 
\end{equation}
in order to get our required thermodynamic functions. Here, the $\mid n\rangle$'s are the eigenvectors and $H_n$ are the eigenvalues. The direct products of the possible vectors $|s_{3/2},m_{3/2}\rangle_i$ for spin-3/2 and $|s_{1/2},m_{1/2}\rangle_j$ for spin-1/2 should be calculated. The direct product, i.e., $|s_{3/2},m_{3/2}\rangle_i\otimes|s_{1/2},m_{1/2}\rangle_j$, gives us eight possible base vectors as 
\begin{eqnarray}
&&|I\rangle=|3/2,3/2\rangle_i|1/2,1/2\rangle_j\,,
\hspace{1.1cm}|II\rangle=|3/2,3/2\rangle_i|1/2,-1/2\rangle_j\,,\nonumber \\
&&|III\rangle=|3/2,-3/2\rangle_i|1/2,1/2\rangle_j\,, 
\hspace{0.45cm}|IV\rangle=|3/2,-3/2\rangle_i|1/2,-1/2\rangle_j\,,\nonumber \\
&&|V\rangle=|3/2,-1/2\rangle_i|1/2,1/2\rangle_j\,,
\hspace{0.7cm}|VI\rangle=|3/2,-1/2\rangle_i|1/2,-1/2\rangle_j\,,\nonumber \\
&&|VII\rangle=|3/2,1/2\rangle_i|1/2,1/2\rangle_j\,,
\hspace{0.6cm}|VIII\rangle=|3/2,1/2\rangle_i|1/2,-1/2\rangle_j.
\end{eqnarray}
These eigenvectors are used to obtain the matrix form of $\hat{\mathcal{H}}_{ij}$ which is $8\times8$ matrix whose elements are obtained from $H_{ij}=\langle i|\hat{\mathcal{H}}_{ij}|j\rangle$ with $i,j=I,II,...,VIII$ and it is found as: 
\begin{equation}
H_{ij}=\left( \begin{array}{cccccccc}
H_{11}&0&0&0&0&0&0&0\\
	 0&H_{22}&-\sqrt{3}*t&0&0&0&0&0\\ 0&-\sqrt{3}*t&H_{33}&0&0&0&0&0\\ 0&0&0&H_{44}&2*t&0&0&0
	\\ 0&0&0&2*t&H_{55}&0&
	0&0\\0&0&0&0&0&H_{66}&-\sqrt{3}*t&0\\ 0&0&0&0&0&-\sqrt{3}*t&H_{77}&0\\ 0&0&0&0&0&0&0&H_{88}
 \end{array}\right) \,,
\end{equation}
where the diagonal matrix elements are found as
\begin{eqnarray}
&&H_{11}=-\frac{3J}{4}-\frac{9D_{i}}{4}-\frac{3h_{i}}{2}-\frac{h_{j}}{2}, \hspace*{0.5cm}
H_{22}=\frac{3J}{4}-\frac{9D_{i}}{4}-\frac{3h_{i}}{2}+\frac{h_{j}}{2}, \nonumber \\&&
H_{33}=-\frac{J}{4}-\frac{D_{i}}{4}-\frac{h_{i}}{2}-\frac{h_{j}}{2},\quad\hspace*{0.65cm}
H_{44}=\frac{J}{4}-\frac{D_{i}}{4}-\frac{h_{i}}{2}+\frac{h_{j}}{2}, \nonumber \\&&
H_{55}=\frac{J}{4}-\frac{D_{i}}{4}-\frac{h_{i}}{2}-\frac{h_{j}}{2},\quad\hspace*{0.85cm}
H_{66}=-\frac{J}{4}-\frac{D_{i}}{4}-\frac{h_{i}}{2}+\frac{h_{j}}{2}, \nonumber \\&&
H_{77}=\frac{3J}{4}-\frac{9D_{i}}{4}-\frac{3h_{i}}{2}-\frac{h_{j}}{2},\hspace*{0.6cm}
H_{88}=-\frac{3J}{4}-\frac{9D_{i}}{4}-\frac{3h_{i}}{2}-\frac{h_{j}}{2},
\end{eqnarray} 
with $t=-{\frac{J}{2}(1-\Delta_{0})}$. The eigenvalues of this matrix $H_{ij}$ are calculated as:
\allowdisplaybreaks
\begin{eqnarray}
&&\lambda_{1}= H_{11},\nonumber \\&&
\lambda_{2}=-{{\sqrt{12t^2 + H_{66}^2-2H_{66}H_{77} +H_{77}^2}-\,H_{66}-H_{77}}\over{2}},\nonumber \\&&
\lambda_{3}={{\sqrt{12t^2 + H_{66}^2-2H_{66}H_{77} +H_{77}^2}+\,H_{66}+H_{77}}\over{2}},\nonumber \\&&
\lambda_{4}=-{{\sqrt{16t^2 + H_{44}^2-2H_{44}H_{55} +H_{55}^2}-\,H_{44}-H_{55}}\over{2}},\nonumber \\&& 
\lambda_{5}={{\sqrt{16t^2 + H_{44}^2-2H_{44}H_{55} +H_{55}^2}+\,H_{44}+H_{55}}\over{2}},\nonumber \\&& 
\lambda_{6}=-{{\sqrt{12t^2 + H_{22}^2-2H_{22}H_{33} +H_{33}^2}-\,H_{22}-H_{33}}\over{2}},\nonumber \\
&&\lambda_{7}={{\sqrt{12t^2 + H_{22}^2-2H_{22}H_{33} +H_{33}^2}+\,H_{22}+H_{33}}\over{2}},\nonumber \\&&  
\lambda_{8}= H_{88}.
\end{eqnarray}
After obtaining the eigenvalues, we are now ready to obtain the partition function which is the main ingredient of our calculation which is found from 
$ Z= \mathrm{Tr}_{ij}[\exp(-\beta H_{ij})]$ and calculated as:\\
\begin{eqnarray}
Z &=&\sum\limits_{n=1}^8 \exp(-\beta \lambda_n) \nonumber \\ &=& \exp(-\beta H_{11})+\exp\left(-\beta H_{88}\right)\nonumber\\
&+& 2\exp\Big[-\beta (H_{22}+ H_{33})\Big]\cosh\left(\frac{\beta \omega_{23}}{2}\right)\nonumber\\
&+& 2\exp\Big[-\beta (H_{44}+ H_{55})\Big]\cosh\left(\frac{\beta \omega_{45}}{2}\right)\nonumber\\
& +& 2\exp\Big[-\beta (H_{66}+ H_{77})\Big]\cosh\left(\frac{\beta \omega_{67}}{2}\right),\nonumber
\end{eqnarray}
where $\beta = \frac{1}{k_\text{B}T}$, $\omega_{23}= \lambda_{3} - \lambda_{2}$, $\omega_{45}= \lambda_{5} - \lambda_{4}$, $\omega_{67}= \lambda_{7} - \lambda_{6}$  and $k_\text{B}$ is the Boltzmann constant.

The magnetizations for spin-3/2 and spin-1/2 are calculated by using the definitions

\begin{equation}
M_{3/2}=\frac{1}{\beta Z} \frac{\partial Z}{\partial h_i} 
\end{equation}
and
\begin{equation}
M_{1/2}=\frac{1}{\beta Z} \frac{\partial Z}{\partial h_j}. 
\end{equation}
In addition, the average magnetization can be defined as
\begin{equation}
M_T=(M_{3/2}+M_{1/2})/2.
\end{equation}
Finally, the free energy of the model can be obtained by using 
\begin{equation}
F=-(1/\beta) \ln Z+J (q-1) M_{3/2} M_{1/2}.
\end{equation}

\begin{figure}[!t]
	\centering
\includegraphics[angle=0,width=0.55\textwidth]{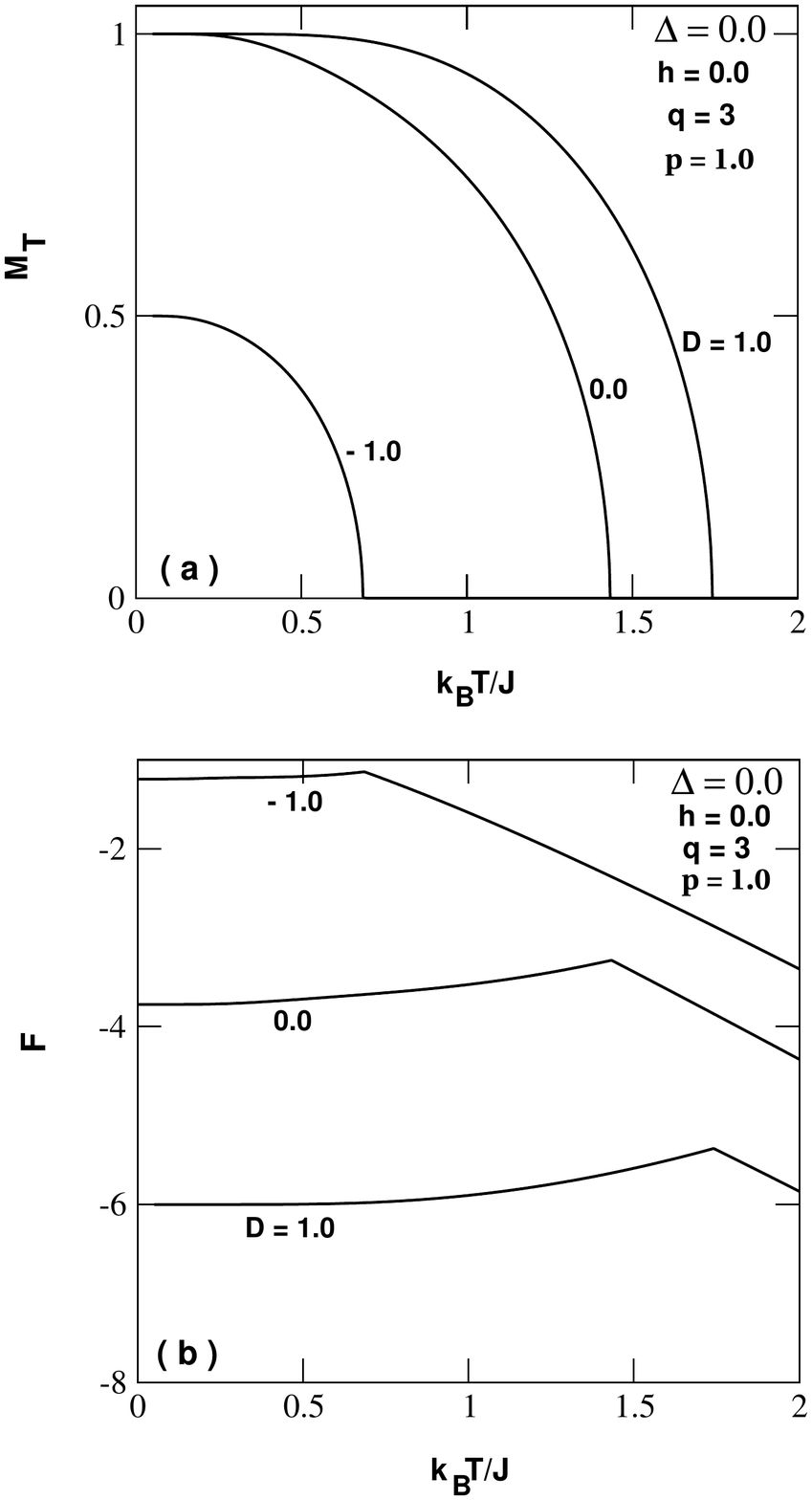}
	\caption{Temperature dependence of global magnetization (panel a) and the free energy (panel b)
of the system when $\Delta=0.0$, $q = 3$, $h = 0.0$ and $p = 1.0$
for selected values of the crystal-field.}
	\label{fig1}
\end{figure}
\begin{figure}[!t]
	\begin{center}
    \includegraphics[angle=0,width=0.48\textwidth]{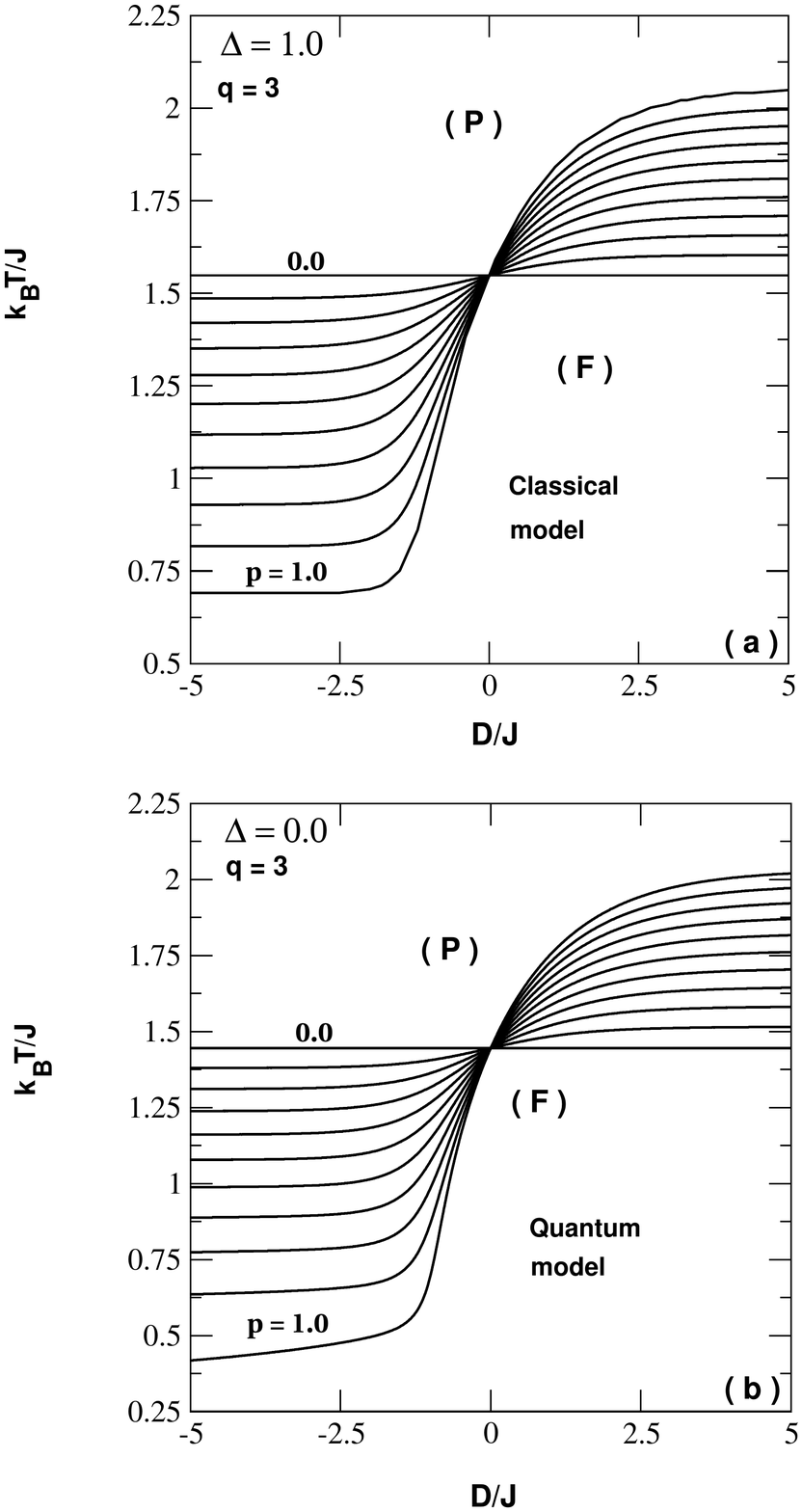}
	\end{center}
	\caption{Phase diagrams of the system in the $(D/J, k_\text{B}T/J)$ plane for selected values 
of the probability $p$ between 0 and 1 with an increment of 0.1 for  $\Delta =1.0$ (panel a) and 
$\Delta = 0.0$ (panel b) and $q = 3$.}
	\label{fig2}
\end{figure}

So far all the equations are obtained for the mixed spin-3/2 and 1/2 HM in the OA. In order to implement the RCF that is given in the bimodal form which turns the crystal field on with probability $p$ and turns it off with $1-p$ 
 \begin{eqnarray}
P(D_i) = p\delta( D_{i}- D) + (1-p)\delta (D_{i})\nonumber
\end{eqnarray}
must be combined with the final forms of magnetizations. Thus, the magnetizations with RCF effects can be obtained from
\begin{eqnarray}
m_{3/2} = \int \frac{1}{\beta Z}\Big[\frac{\partial{Z}}{\partial{ h_{i}}}\Big]P(D_{i})\rd D_{i}
\end{eqnarray}
and
\begin{eqnarray}
m_{1/2}=\int \frac{1}{\beta Z}\Big[\frac{\partial{Z}}{\partial{ h_{j}}}\Big]P(D_{i})\rd D_{i}.
\end{eqnarray}
Lastly, the free energy with RCF is also given as 
\begin{eqnarray}
f&=& -\frac{1}{\beta}\int \ln Z P(D_{i})\rd D_{i}.
\end{eqnarray}
In the next section, we are going to illustrate our findings in terms of possible phase diagrams,
thermal variations of magnetizations and magnetic hysteresis loops. 

\section{The phase diagrams}

 We  start this section by
analyzing the thermal behavior of the order parameters which are reduced in the model to the global
magnetization of the  system. This is depicted in figure~\ref{fig1}a for the coordination number $q=3$. As usually 
observed, the total magnetization starts from its saturation values and, as thermal fluctuations grow in
the system, it  continuously decreases and finally vanishes at the critical temperature $T_c$. As it can 
be observed, the displayed curves  correspond to selected values of the crystal-field $D$. It immediately appears
that when $D$ increases, a coexistence point between thermodynamic phases (1/2, 1/2) and (3/2, 1/2) may appear 
because there exist  two different saturation values of the global magnetization. The transition which is previously observed when $M$ vanishes is of the second-order and $T_c$ 
increases when the value of $D$ is raised. This means that $D$ produces some stabilizing effect on the
ordered ferrimagnetic phase whose volume also increases. In figure~\ref{fig1}b, the free energy is illustrated and shows 
two  different behaviors separated by a peak which is associated with the critical temperature~$T_c$. One can
see the first region where it slowly increases. This increase becomes pronounced when $D$ gets larger values.
In the second region, a very fast decay of the free energy is observed because the entropy term exceeds the 
internal energy leading to  almost straight lines. 
Similar 
results were reported in~\cite{Guli} which dealt with
 thermodynamic properties of a spin-1/2 Heisenberg 
ferromagnetic system within the same Oguchi's approximation.
 Therein, the internal energy $U$ is monitored and
 exhibited, as the  entropy, a constant for $ T > T_c$.
\begin{figure}[!t]
	\begin{center}
		\includegraphics[angle=0,width=0.55\textwidth]{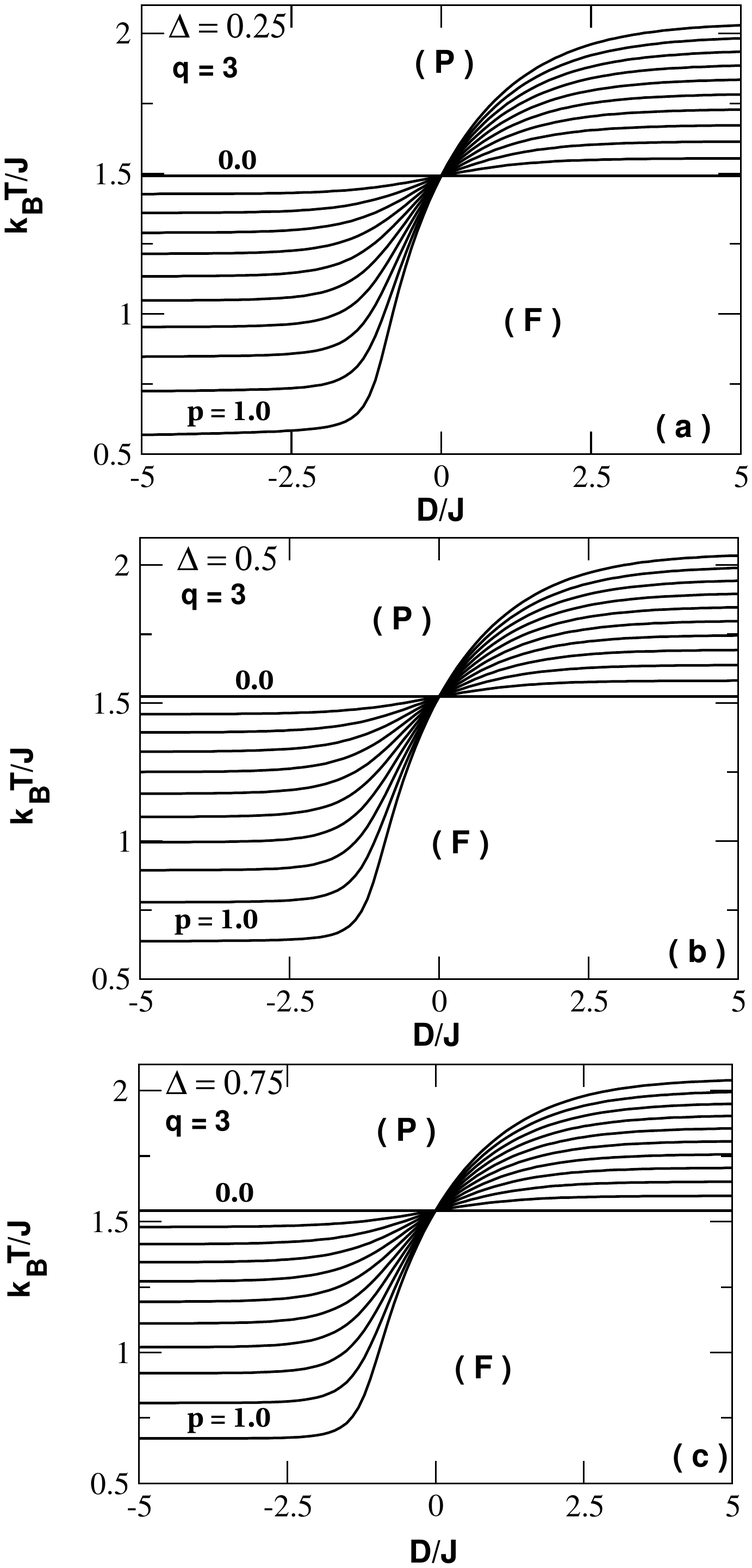}
	\end{center}
	\caption{Phase diagrams of the system in the $(D/J, k_\text{B}T/J)$ plane for selected values
of the probability $p$ between 0 and 1 with an increment of 0.1 for  $\Delta = 0.25$  (panel a); 
$\Delta = 0.5$ (panel b) and  $\Delta = 0.75$ (panel c) and $ q = 3$.}
	\label{fig3}
\end{figure}
For varying values of the crystal-field  interaction parameter $D$ and probability distribution $p$, one is
able to devise a phase diagram for the model in the  ($D/J,k_\text{B}T/J$) plane. This is done in figures~\ref{fig2}a and b for the
classical version ($\Delta=1$) and the quantum version ($\Delta=0$). Both panels show similar qualitative trends. 
Some interesting features emerge. In particular, it is observed that the critical temperature $T_c$  becomes 
insensitive to the value of the crystal-field parameter when this value becomes relatively high at any value
of the probability $p$. This behavior is observed in several previous works. In fact, \cite{Dak} that 
dealt with exact calculations on the mixed spin-1/2 and spin-S model on the square lattice reported similar trends.
For large and positive/negative values of~$D$, thermodynamic phases (3/2,1/2)/(1/2,1/2) should prevail in the system. 
It could be observed in the case that the difference between the asymptotic values of the critical temperature 
$T_c$ is a decreasing function of the parameter $p$ that vanishes at $p=0$. Our calculations for these asymptotic
values lead to the following results. In the classical model, for $p=0;\,0.5;\,1$ and $D\rightarrow  -\infty/+\infty$,
we get respectively $k_\text{B}T/J=1.548/1.548$; $k_\text{B}T/J=1.200/1.810$; $k_\text{B}T/J=0.691/2.051$. In the quantum model case, 
for $p=0;\,0.5;\,1$ and $D\rightarrow  -\infty/+\infty$, one respectively obtains  $k_\text{B}T/J=1.446/1.446$; 
$k_\text{B}T/J=1.075/1.769$; $k_\text{B}T/J=0.346/2.037$. It emerges that the change of the gap when $p$ runs from $0$ to $1$ becomes considerably larger in the case of negative values of $D$ than for positive values in both models. When
comparing the results displayed in both panels, it evidently appears that the quantitative change of the critical
temperature $T_c$ could be clearly seen in the negative $ D $-range considered whereas only similar qualitative 
trends are observed beyond. As an example,  for $p=1$, the critical temperature $T_c$ for the quantum model 
is about half of that of the classical model, which means that quantum effects play a key role 
in the magnetic properties of the quantum system for negative values of the  parameter 
$D$ which are associated with low values of $T_c$. For this value of $p$, quantum effects 
on the critical temperature are negligible at high values of $D$ which induce $T_c$.

After studying the phase diagrams for the two extreme cases, $\Delta=0;1$, the anisotropic
case $\Delta \neq 0;1$ needs to be somewhat clarified. Results concerning this intermediate 
case are illustrated in figure~\ref{fig3}. As it can be seen from different panels, the anisotropic case
shows the same qualitative behavior of phase boundaries with the extreme quantum case.
At fixed values of $p$, one sees that $T_c$ is an increasing function of $\Delta$.
The value corresponding to the saturation of $T_c$ in the negative $ D $-range also increases
when the value of $\Delta$ is raised.

\begin{figure}[!t]
	\begin{center}
		\includegraphics[angle=0,width=0.6\textwidth]{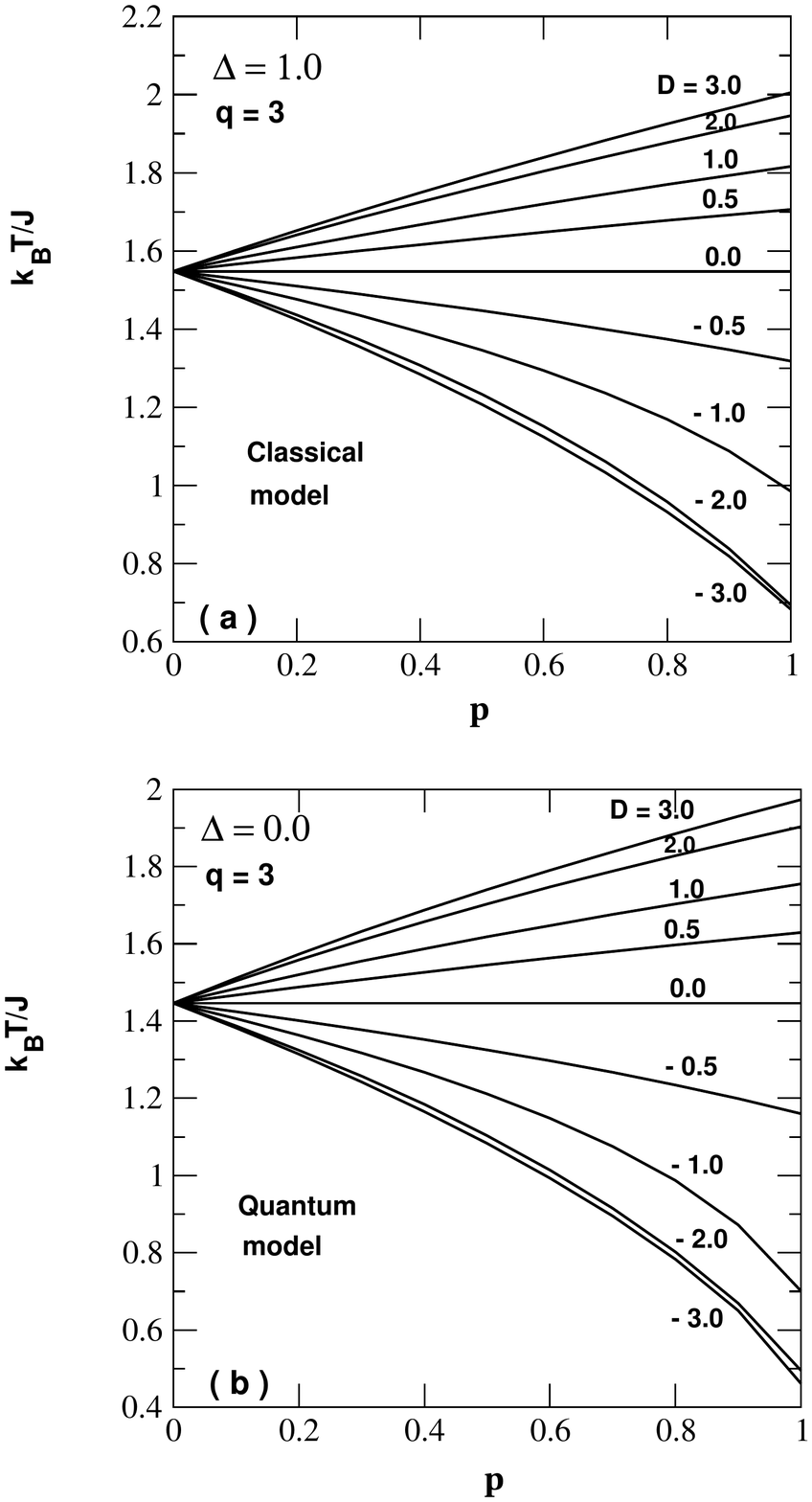}
	\end{center}
	\caption{Phase diagrams of the system in the $(p, k_\text{B}T/J)$ plane for selected  values of
 $D$ for $\Delta =1.0$ (panel~a); $\Delta = 0.0$ (panel b) and $q = 3$. }
	\label{fig4}
\end{figure}

In figure~\ref{fig4}, we show another way to present the results
reported in figure~\ref{fig2}.  We display  phase diagrams in the ($p,k_\text{B}T/J$) plane for selected 
values of the crystal-field for the classical  and quantum models. For $\Delta=0$, one gets
a straight and horizontal second-order transition line. For increasing values of $D$, we get
 increasing values of $T_c$  whereas for decreasing negative values, the contrary holds.
Throughout the previous figures, our numerical calculations only show second-order transition.
As it is shown in the following, one can get first-order transitions in the model at a relatively 
high value of the coordination parameter $q$ as well as in  the presence of an external magnetic field
constraint.

\begin{figure}[!t]
	\begin{center}
    \includegraphics[angle=0,width=0.55\textwidth]{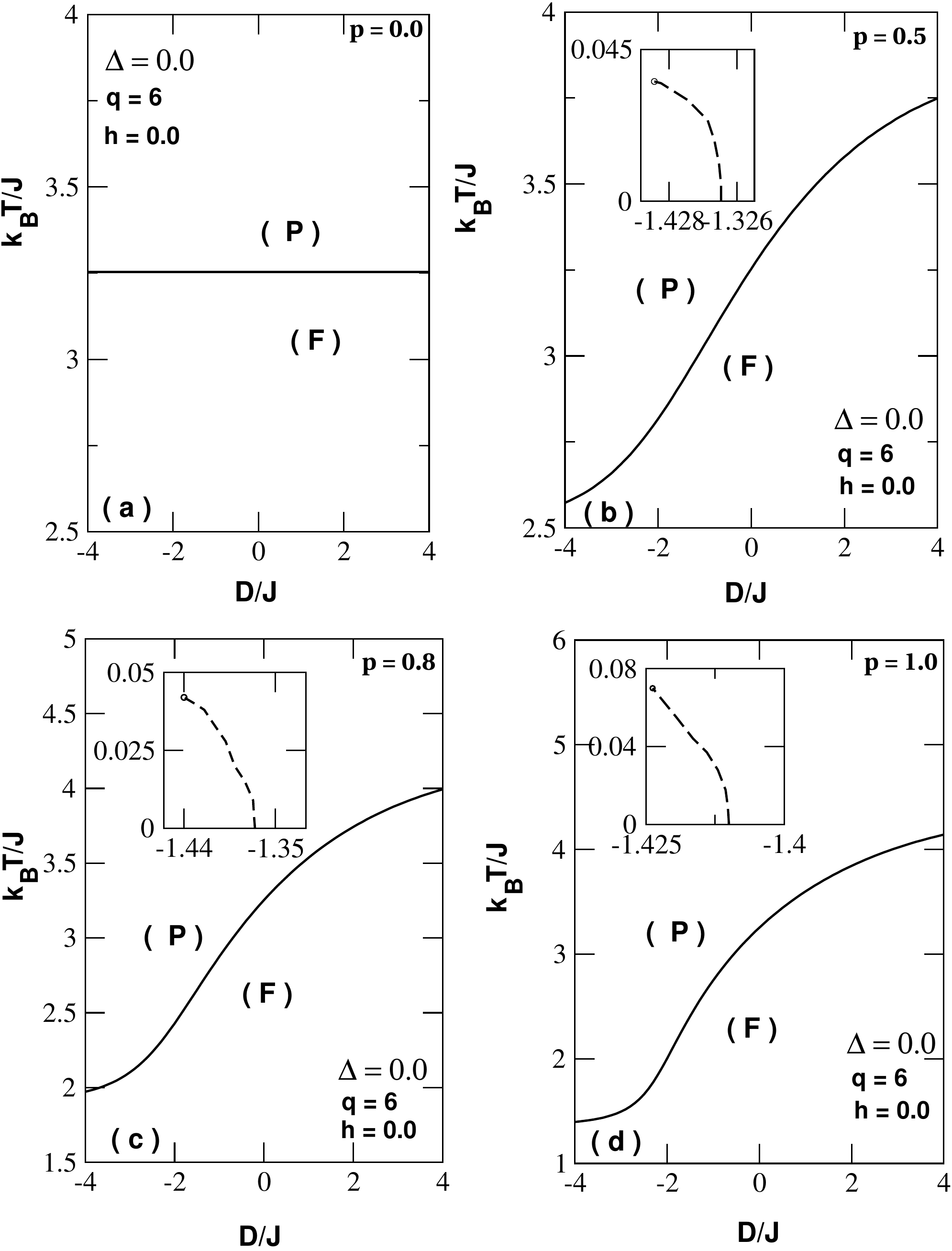}
	\end{center}
	\caption{Phase diagrams of the system in the  $(D/J, k_\text{B}T/J)$ plane for
 $p = 0.0$ (panel a); $ p = 0.5$ (panel b); $p = 0.8$ (panel c); $p = 1.0$ (panel d) and $ q = 6$.
 A first-order transition phenomenon is detected at low temperature for negative values of $D/J$.}
	\label{fig5}
\end{figure}

In fact, in figure~\ref{fig5}, we present a set of phase diagrams for the quantum model in the $(D/J,k_\text{B}TJ)$
plane for selected values of the probability $p$ and coordination number $q=6$. The case $q=6$ 
should correspond to the simple cubic lattice case. The number of nearest-neighbor interactions
is then increased. For $p=0$, one gets the same straight line separating the ferrimagnetic and 
the disordered paramagnetic phases as in figures~\ref{fig2},\ref{fig3}. For $p\geqslant0$, the results indicate the appearance
of  first-order transitions that are localized by jumps in the global magnetization of the system.
This is observed in the very low-temperature range and for a $D/J$ value around $-1.4$. Thus, the results 
for $(p=1; \Delta=0)$ (panel d) correspond to the results reported in  figure~\ref{fig3} of \cite{Bobak3}. 
Let us remind the reader that the introduction of the probability distribution on the crystal-field 
is the only difference between our model and that of  \cite{Bobak3}.  A careful observation of
the onset of the first order transition when the parameter $p$ varies shows that the value of $D$
associated with this onset decreases with $p$. Our calculations show that the first order transition
lines observed in the panels terminate at
end-points. Hence, the critical values associated with $p=0.5,\,0.8,\,1$ are
respectively: $(D/J=-1.45,k_\text{B}T/J=0.035)$, $(D/J=-1.44,k_\text{B}T/J=0.042)$, $(D/J=-1.424,k_\text{B}T/J=0.071)$.

\begin{figure}[!t]
	\begin{center}
    \includegraphics[angle=0,width=0.53\textwidth]{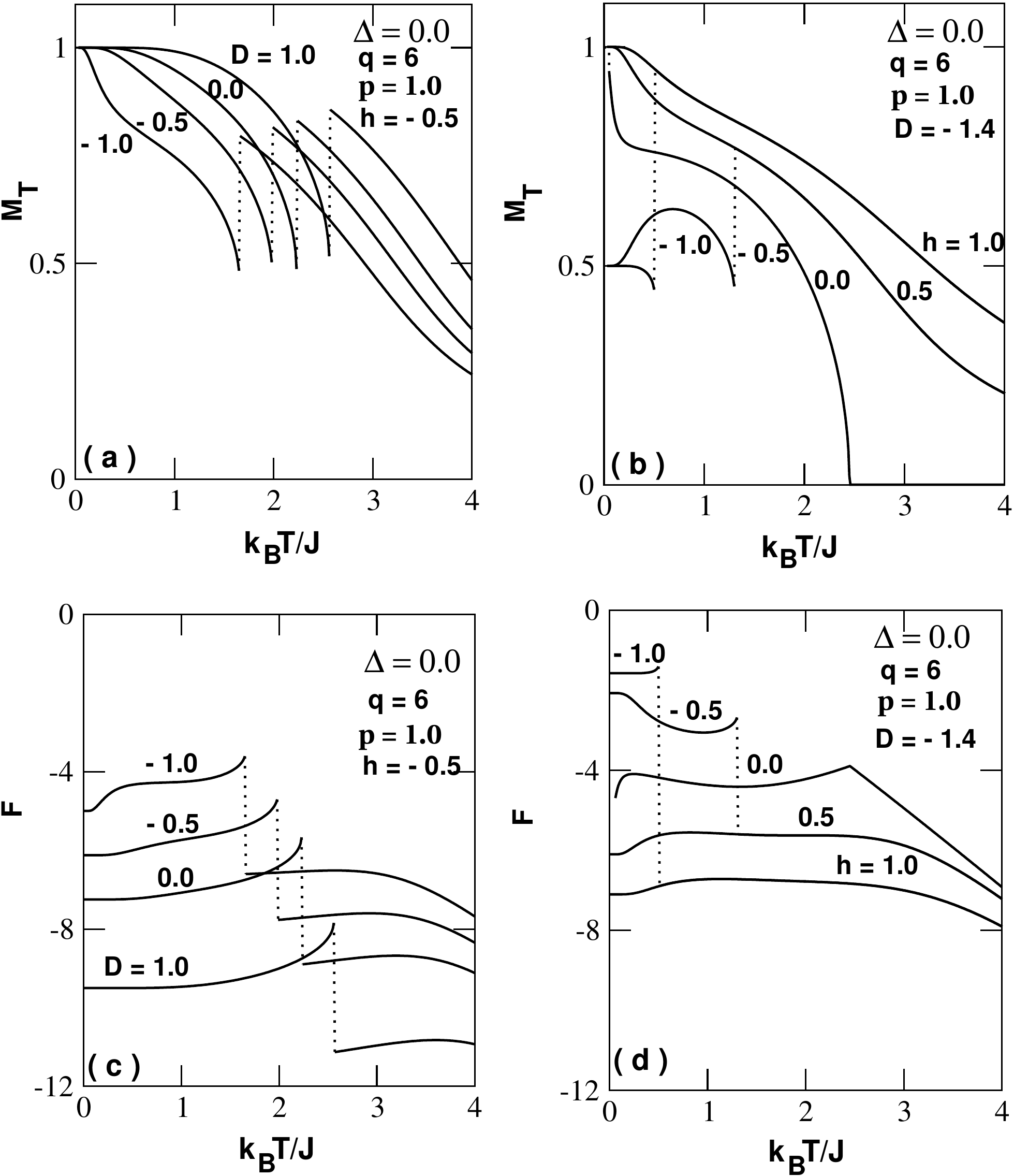}
	\end{center}
	\caption{ The behaviors of the magnetization ( panels a,b) and the free energy ( panels c,d) as functions of 
 the temperature $T$. In panel a, the effect of the crystal-field on the
 first-order transition is illustrated whereas  panel b shows
  how the values of h displace this transition.}
	\label{fig6}
\end{figure}

In what follows, the influence of the applied magnetic field is
examined on the quantum system. Figure~\ref{fig6}a illustrates 
the change of the first-order transition temperature with the value of the crystal-field $D$.
At a constant value $h=-0.5$,
one sees that this temperature increases when the value of the crystal-field increases. 
 The corresponding free energies
 in figure~6c show discontinuities at the transition temperatures.
 This insures that we are in  fact in the presence of true
 first-order transitions rather than the jumps of the system from 
 metastable states to stable (steady) states when thermal 
 fluctuations are enhanced. Figure~\ref{fig6}b illustrates how the first-order 
 transition is affected when $h$ varies. It could be observed 
 that no transition appears with positive values of the field $h$.
 All these observed behaviors result 
 from the competition between different model parameters of the system when the temperature varies.

\begin{figure}[!t]
	\begin{center}
    \includegraphics[angle=0,width=0.49\textwidth]{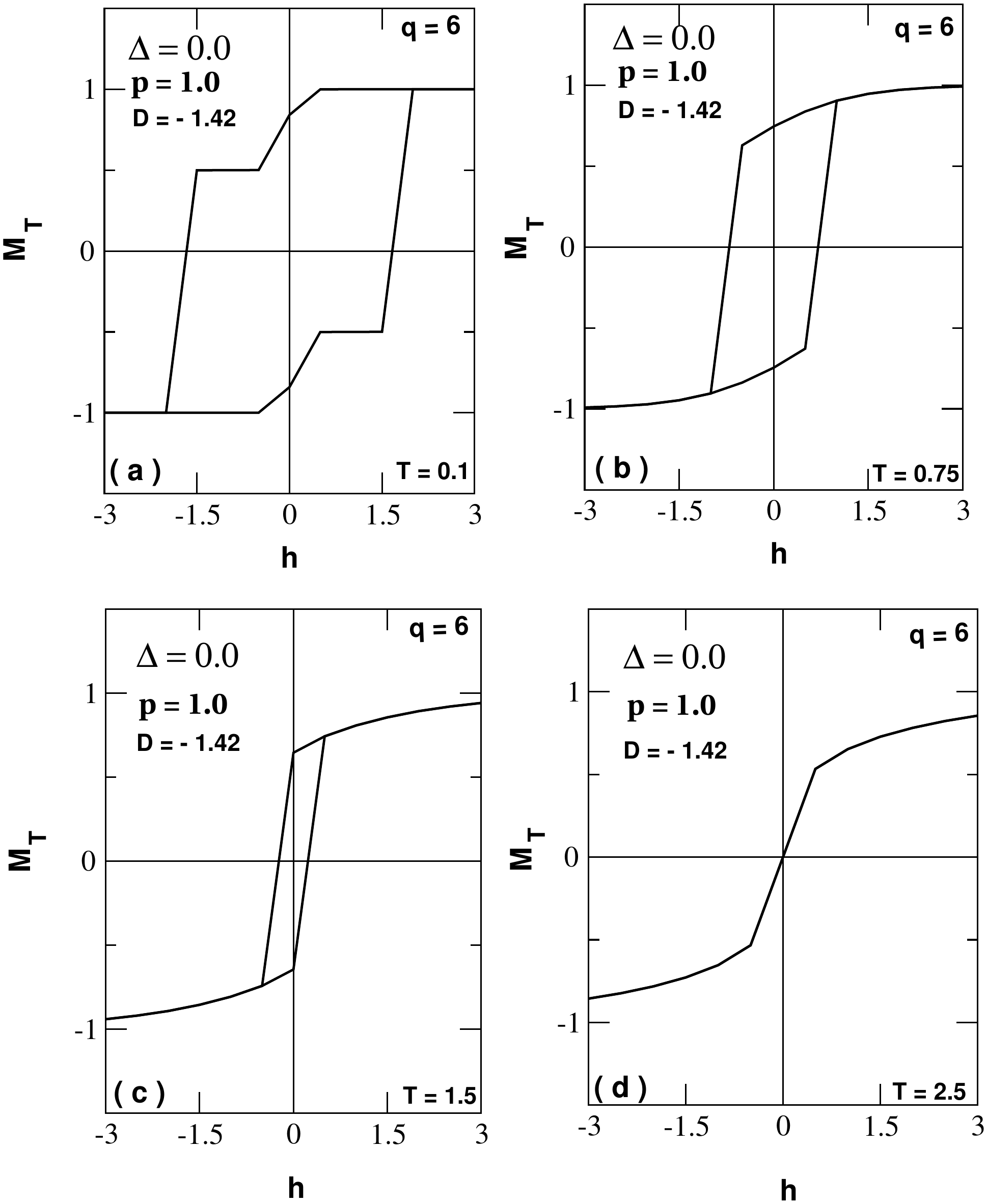}
	\end{center}
	\caption{Typical hysteresis behavior of the system as a function of the 
 applied field h for selected values of the temperature $T$.}
	\label{fig7}
\end{figure}

 Different panels of figure~\ref{fig7} 
 show hysteresis phenomena and how the system disorders when the temperature is raised from zero.
  It also appears that the coercitive field $h_c$ and the remannent magnetization $m_r$ 
  decrease with the temperature $ T $
   since at high temperature a critical hysteresis is obtained.

\begin{figure}[!t]
	\begin{center}
    \includegraphics[angle=0,width=0.6\textwidth]{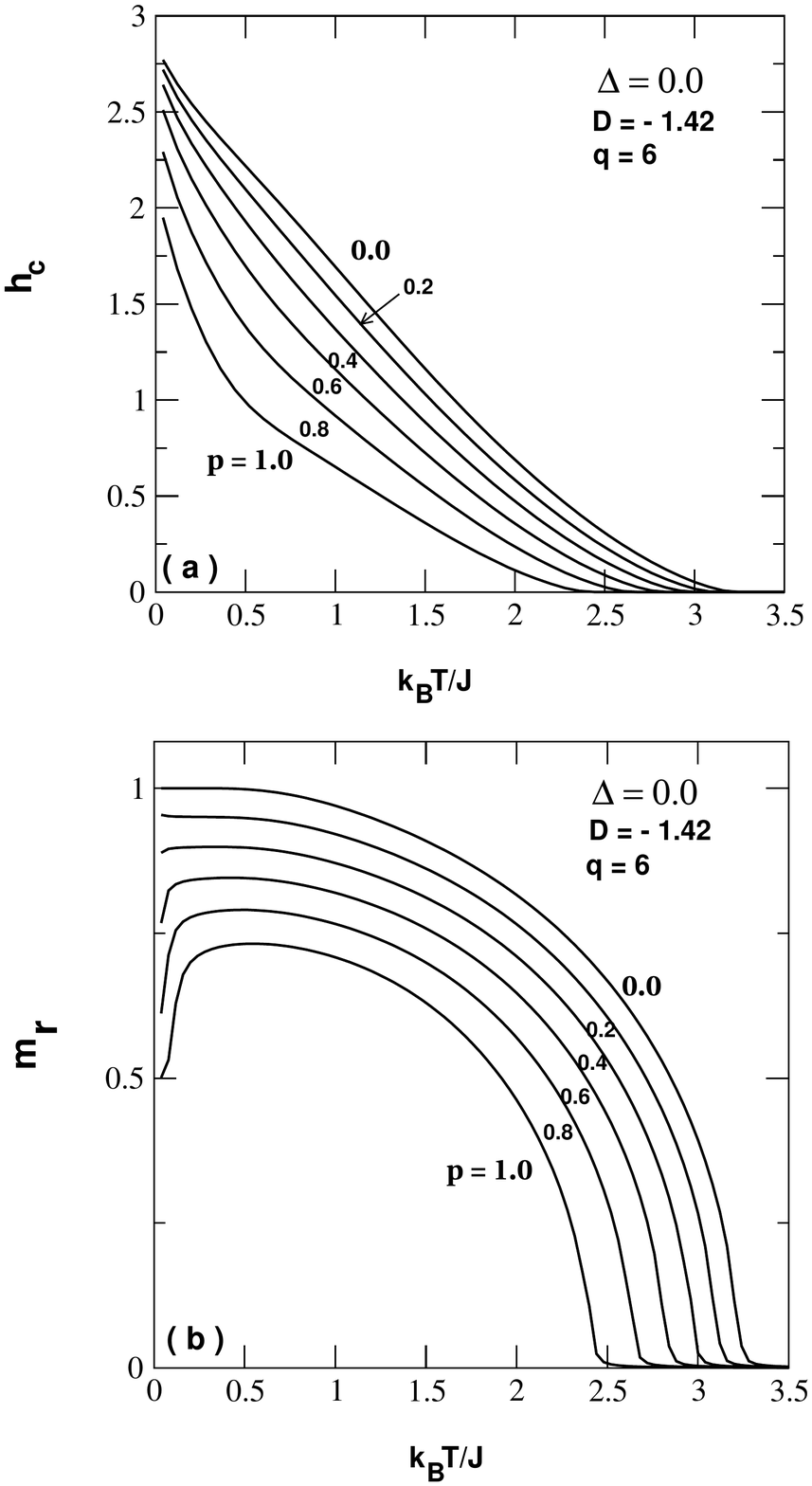}
	\end{center}
	\caption{The coercive field $h_c$ (panel a) and the remanent magnetization (panel b)
 as functions of the temperature for selected values of the model parameters.}
	\label{fig8}
\end{figure}

   Figure~\ref{fig8} gives detailed illustrations of that.
   Ou results for $h_c$ look similar to those displayed in  \cite{mounirou,bati}.
   The remanent magnetization shows a 
   narrow low-temperature region where it increases with $ T $.
   Beyond this region, it behaves 
   like the global magnetization and vanishes at  critical
   temperatures corresponding to selected values of $p$.
   A similar behavior is also observed for the coercivity.
   \newpage

\section{Brief summary and conclusion}
 
          In this work, the mixed spin-3/2 and spin-1/2 Heisenberg model was investigated
	  by means of the Oguchi approximation which is a kind of a Mean Field treatment. 
	  The classical, quantum and anisotropic cases
	  were analyzed in the presence of a random crystal-field which was set on the system
	  with probability $p$ or turned off with probability $1-p$. Computed magnetizations
	  in the quantum case  showed the presence of second-order phase transitions for
	   the coordination number $q=3$ in the absence of the external magnetic field $ h $. The
	    free energy $F$ was found to increase with thermal fluctuations in the ferrimagnetic phase
	    whereas a decrease  emerged at the onset of total disorder in the system.
	    For $q=3$, several phase diagrams are plotted for varying values of the 
	    anisotropy $\Delta$  and probability $p$.  It was observed that quantum
	     effects are almost absent in the positive $ D $-range whereas for relatively low and 
	     negative
	      values of $ D $ they appeared important, in particular in the quantum model for $p=1$.
	      Our calculations did not show first-order transitions for $q=3$. Instead, for $q=6$,
	       this phenomenon appeared at a very low temperature in the negative
	        $ D $-range even in the absence of the external field constraint. It was also revealed
		 that for positive values of $h$, the first-order transition was absent whereas beyond,
		  at least in the range of selected values of $h$ used in the computations, it 
		   appeared and the corresponding temperature decreases with the absolute value of $h$.
		  The existence of first-order transition induced the presence of  a hysteresis 
		  phenomenon in the quantum model. It was also shown that thermal fluctuations
		  had a dramatic effect on the size of the hysteresis 
		  loop which decreases as the temperature increases
		  and finally generates a critical hysteresis.
		  Finally, we found that coercivity and remanence 
		  decrease with the temperature in a wide range of the temperature $ T $.

\newpage
\ukrainianpart

\title{Анізотропна модель Гайзенберга змішаного типу спін-3/2 та спін-1/2 під дією випадкового кристалічного поля}
	
\author{Д. Сабі Таку\refaddr{label1,label3}, M. Каріму\refaddr{label2,label3}, Ф. Гонтінфінде\refaddr{label3,label4},  
	Е. Албайрак\refaddr{label5}}
		
		\addresses{
		\addr{label1} Вища політехнічна школа Абомей-Калаві (EPAC-UAC), факультет природничих наук, Республіка Бенін
		\addr{label2} Національна вища школа енергетики і процесів (ENSGEP), Абомей, Республіка Бенін  
		\addr{label3} Інститут матаматичних і фізичних наук (IMSP),  Республіка Бенін  
		\addr{label4} Університет Абомей-Калаві, фізичний факультет, Республіка Бенін
		\addr{label5} Університет Ерджієс, фізичний факультет, 38039, Кайсері, Туреччина}
		
\makeukrtitle

	\begin{abstract}
		Термодинамічні властивості  змішаної спін-3/2 та спін-1/2 моделі
	 Гайзенберга досліджується в наближенні Огучі
		за наявності випадкового кристалічного поля (ВКП).
		ВКП або вводиться з імовірністю $ p $, або вимикається з
		ймовірністю $ 1-p $ випадково. Досліджуються теплові коливання глобального намагнічування
		та вільна енергія системи  для того, щоб побудувати фазові діаграми
		для класичних, квантових та анізотропних випадків. Отримані різні результати показують, що між ними не існує якісних змін. Встановлено  присутність квантових ефектів
	 у квантовій моделі в негативному $ D $-діапазоні. Цей феномен виявляє
	 сильний зменшувальний вплив на критичну температуру, яка стає набагато нижчою, ніж
		у класичному випадку. При наявності зовнішнього
		поля, спостерігалось, що коерцитивність і залишкова намагніченість зменшуються в широкому діапазоні
		абсолютних температур. 
		
\keywords обмінна анізотропія, модель Гайзенберга, наближення Огучі, змішаний спін,  кристалічне поле

\end{abstract}


\begin{thebibliography}{99}
\bibitem{Sousa}  Ricardo de Sousa J.,  Lacerda F.,   Fittipaldi I.P., Physica A, 1998, {\bf{258}}, 221--229, \\\doi{10.1016/S0378-4371(97)00537-2}.
\bibitem{Mert}  Mert G., J. Magn. Magn. Mater., 2015, {\bf{394}}, 126--129, \doi{10.1016/j.jmmm.2015.06.062}.	
\bibitem{Bobak}  Bob\'{a}k A.,  Pokorn\'{y} V.,   Dely J., Physica A,  2009, {\bf{388}}, 2157--2167, \doi{10.1016/j.physa.2009.02.017}.	
\bibitem{Bobak1} 	Bob\'{a}k A.,  Dely J.,  \v{Z}ukovi\v{c} M., Physica A, 2011, {\bf{390}},   1953--1960, \doi{10.1016/j.physa.2011.02.028}.
\bibitem{Bobak2}  Bob\'{a}k A.,  Pokorny V.,   Dely J., J. Phys. Conf. Ser., 2010, {\bf{200}}, 022001, \doi{10.1088/1742-6596/200/2/022001}.
\bibitem{Albayrak0}  Albayrak E., J. Supercond. Novel Magn., 2017, {\bf{30}} 2555--2561, \doi{10.1007/s10948-017-4079-4}.
\bibitem{Albayrak1}  Albayrak E., Physica A, 2017, {\bf{486}}, 161--167, \doi{10.1016/j.physa.2017.05.042}.
\bibitem{Bobak3} 	Bob\'{a}k A.,  Feckov\'{a} Z.,   \v{Z}ukovi\v{c} M., J. Magn. Magn. Mater., 2011, {\bf{323}}, 813--818,\\ \doi{10.1016/j.jmmm.2010.11.003}.	 


\bibitem{Kaneyoshi0}  Kaneyoshi T.,  Ja\v{s}\v{c}ur M.,   Tomczak P., J. Phys: Condens. Matter,  1992, {\bf{4}}, L653--L658,\\ \doi{10.1088/0953-8984/4/49/002}.
\bibitem{Kaneyoshi2}  Kaneyoshi T., Physica A,  1995, {\bf{215}}, 378--386, \doi{10.1016/0378-4371(94)00306-E}.
\bibitem{Wei}  Wei G.Z.,  Liang Y.Q.,  Zhang Q.,   Xin Z.H., J. Magn. Magn. Mater.,  2004, {\bf{271}}, 246--253, \\\doi{10.1016/j.jmmm.2003.09.043}.
\bibitem{Essaoudi}  Essaoudi I., B\"{a}rner K.,  Ainane A.,   Saber M., Physica A, 2007, {\bf{385}}, 208--220, \doi{10.1016/j.physa.2007.06.037}.
\bibitem{Albayrakx}  Yigit A.,   Albayrak E., J. Magn. Magn. Mater.,  2013, {\bf{329}}, 125--128, \doi{10.1016/j.jmmm.2012.10.011}.
\bibitem{Jiang1}  Jiang W.,  Wei G.Z.,   Du A., J. Magn. Magn. Mater., 2002, {\bf{250}},  49--56, \doi{10.1016/S0304-8853(02)00351-7}.
\bibitem{Saber}  Htoutou K.,  Ainane A.,   Saber M., J. Magn. Magn. Mater.,  2004, {\bf{269}}, 245--258, \\ \doi{10.1016/j.jmmm.2003.07.002}.
\bibitem{Liang}  Liang Y.Q.,  Wei G.Z.,  Zhang Q.,  Wei Q.,   Zang S.L., Chin. Phys.,  2004, {\bf{13}}, 2147, \\ \doi{10.1088/1009-1963/13/12/030}.
\bibitem{Liang1}  Liang Y.Q.,  Wei G.Z.,  Song G.L., Phys. Status Solidi B, 2004, {\bf{241}}, 1916, 1916--1922, \\ \doi{10.1002/pssb.200301996}.

 
\bibitem{Kaneyoshi1}  Kaneyoshi T., J. Magn. Magn. Mater., 1995, {\bf{151}}, 45--53, \doi{10.1016/0304-8853(95)00403-3}.
\bibitem{Bobak0}  Bob\'{a}k A.,   Jur\v{c}i\v{s}in M., J. Magn. Magn. Mater., 1996, {\bf{163}}, 292--298, \doi{10.1016/S0304-8853(96)00353-8}. 
\bibitem{Liang2}  Liang Y.Q.,  Wei G.Z.,   Song G.L., Phys. Status Solidi B, 2008, {\bf{245}}, 2586, \doi{10.1002/pssb.200844136}.
\bibitem{Benayad0}  Benayad N.,  Dakhama A.,  Kl\"{u}mper A.,   Zittartz J., Z. Phys. B: Condens. Matter, 1996, {\bf{101}}, 623--630, \doi{10.1007/s002570050255}.
\bibitem{Jiang0}  Jiang W.,  Wei G.Z.,   Xin Z.H., Physica A, 2001, {\bf{293}}, 455--464, \doi{10.1016/S0378-4371(01)00008-5}.
\bibitem{Bobak1x}  Bob\'{a}k  A.,  Horv\'{a}th D., Phys. Status Solidi B, 1999, {\bf{213}}, 459, \\ \doi{10.1002/(SICI)1521-3951(199906)213:2<459::AID-PSSB459>3.0.CO;2-0}.


\bibitem{xBuendia}  Buendia G.M.,   Cardano R., Phys. Rev. B, 1999, {\bf{59}}, 6784, \doi{10.1103/PhysRevB.59.6784}. 
\bibitem{xAlbayrak}  Albayrak E.,   Al\c{c}{\i} A., Physica A, 2005, {\bf{345}}, 48--60, \doi{10.1016/j.physa.2004.04.134}.
\bibitem{xStrecka}  Stre\v{c}ka J.,   \v{C}anov\'{a} L., Condens. Matter Phys., 2006, {\bf{9}}, 179, \doi{10.5488/CMP.9.1.179}.
\bibitem{xJascur}  Ja\v{s}\v{c}ur M.,   Stre\v{c}ka J., Physica A, 2005, {\bf{358}}, 393, \doi{10.1016/j.physa.2005.07.010}.
\bibitem{xZhang}  Zhang X.,   Kong X.M., Physica A, 2006, {\bf{369}}, 589--598, \doi{10.1016/j.physa.2006.02.014}.
\bibitem{xStrecka0}  Stre\v{c}ka J., Phys. Status Solidi B, 2006, {\bf{243}}, 708, \doi{10.1002/pssb.200642018}.
\bibitem{xKissova}  Ki\v{s}\v{s}ova J.,   Stre\v{c}ka J., Acta Phys. Pol. A, 2008, {\bf{113}}, 445--448, \doi{10.12693/APhysPolA.113.445}. 
\bibitem{xBahmad0}  Bahmad L.,  Benayad M.R.,  Benyoussef A.,   El Kenz A., Acta Phys. Pol. A, 2011, {\bf{119}}, \\ 740--746, \doi{10.12693/APhysPolA.119.740}.
\bibitem{xAlbayrak2}  Albayrak E., Chin. Phys. B, 2012, {\bf{21}}, 067501, \doi{10.1088/1674-1056/21/2/020511}.
\bibitem{xAlbayrak3}  Albayrak E.,   Bulut T., J. Magn. Magn. Mater., 2007,  {\bf{316}}, 81--89, \doi{10.1016/j.jmmm.2007.03.202}.
\bibitem{xAlbayrak4}  Albayrak E., Physica B, 2016, {\bf{494}}, 91--95, \doi{10.1016/j.physb.2016.04.028}.
\bibitem{Guli}  Mert G., J. Magn. Magn. Mater.,  2015, {\bf{394}}, 126--129, \doi{10.1016/j.jmmm.2015.06.062}.
\bibitem{Dak} Dakhama A.,  Azhari M.,   Benayad N., J. Phys. Commun., 2018,  {\bf{2}}, 065011, \doi{10.1088/2399-6528/aacbbe}.
\bibitem{mounirou} Karimou M.,  Yessoufou R.A.,  Ngantso G.D.,  Hontinfinde F.,   Benyoussef A., J. Supercond. Novel Magn.,
	 2019, {\bf{32}}, 1769--1779, \doi{10.1007/s10948-018-4876-4}.
\bibitem{bati} Bati M.,   Erta\c{s} M., J. Supercond. Novel Magn., 2016, {\bf{29}}, 2835--2841, \doi{10.1007/s10948-016-3620-1}.
\end{thebibliography}
\end{document}